\documentclass[nofootinbib,twocolumn,showpacs,preprintnumbers,pre,aps]{revtex4-1}

\usepackage{graphicx}
\usepackage{bm}
\usepackage{amsmath}
\usepackage{amssymb}
\usepackage{color}

\begin{document}

\title{Swimmer-microrheology}%

\author{Kento Yasuda}

\author{Ryuichi Okamoto}

\author{Shigeyuki Komura}\email{komura@tmu.ac.jp}

\affiliation{
Department of Chemistry, Graduate School of Science and Engineering,
Tokyo Metropolitan University, Tokyo 192-0397, Japan}

\date{\today}

\begin{abstract}
We discuss a locomotion of a three-sphere microswimmer in a viscoelastic medium
and propose a new type of active microrheology.
We derive a relation which connects average swimming velocity and frequency-dependent 
viscosity of the surrounding medium.
In this relation, the viscous contribution can exist only when the time-reversal symmetry is broken, 
whereas the elastic contribution is present only when the structural symmetry of the swimmer is broken. 
The Purcell's scallop theorem breaks down for a three-sphere swimmer in a 
viscoelastic medium.
\end{abstract}

\maketitle

Microrheology is one of the most useful techniques to measure rheological properties 
of soft matter and various biological materials including cells~\cite{SM10,Chen10}.
There are two different methods;  passive microrheology and active microrheology.
In the passive microrheology, both local and bulk mechanical properties of a medium 
can be extracted from a Brownian motion of a probe particle~\cite{MW95,Mason00}.
In this method, the generalized Stokes-Einstein relation (GSER) is used to analyze 
thermal diffusive motions.
In the active microrheology, on the other hand, the probe is actively pulled through the
fluid, with the aim of driving the medium out-of-equilibrium and measuring mechanical 
responses~\cite{GSOMS,Schnurr97}. 
Within the linear response theory, the generalized Stokes relation (GSR) is employed to 
obtain the frequency-dependent complex shear modulus.

\begin{figure}[b]
\begin{center}
\includegraphics[scale=0.35]{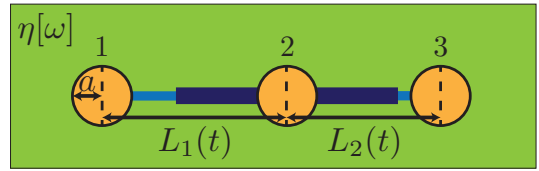}
\end{center}
\caption{
The Najafi-Golestanian's three-sphere swimmer model.
Three identical spheres of radius $a$ are connected by arms of lengths $L_1(t)$ and  $L_2(t)$
and undergo time-dependent cyclic motions.
The swimmer is embedded in a viscoelastic medium characterized by a frequency-dependent 
complex shear viscosity $\eta[\omega]$. 
}
\label{model}
\end{figure}

In this Letter, we propose a new type of active microrheology using a microswimmer.
Microswimmers are tiny machines that swim in a fluid like sperm cells or motile bacteria, 
and are expected to be applied to microfluidics or 
microsystems~\cite{Lauga092}.
As one of the simplest microswimmers, we consider Najafi-Golestanian's 
three-sphere swimmer model~\cite{Golestanian04,Golestanian08}, where three in-line 
spheres are linked by two arms of varying length (see Fig.~\ref{model}).
Recently such a swimmer has been experimentally realized~\cite{Grosjean}.
We investigate its motion in a general viscoelastic medium, and obtain a relation which 
connects the average swimming velocity and the frequency-dependent complex 
shear viscosity of the surrounding viscoelastic medium.
We show explicitly that the absence of the time-reversal symmetry of the swimmer 
motion leads to the real part of the viscosity, whereas the absence of the structural 
symmetry of the swimmer is reflected in the imaginary part of the viscosity.
Hence we call it as ``swimmer-microrheology". 
Our result also indicates that the Purcell's scallop theorem~\cite{Purcell77,Lauga11}, 
which states that time-reversible body motion cannot be used for locomotion in a 
Newtonian fluid, breaks down 
for a three-sphere swimmer 
in viscoelastic media if the structural symmetry is violated.

The general equation that describes the hydrodynamics  of low Reynolds number flow in  
a viscoelastic medium is given by the following generalized Stokes equation~\cite{Granek11}:
\begin{equation}
0=
\int_{-\infty}^t {\rm d}t'\,\eta(t-t') \nabla^2 \mathbf{v}(\mathbf{r},t') -\nabla p(\mathbf{r},t).
\label{StorksEve}
\end{equation}
Here $\eta(t)$ is the time-dependent shear viscosity, $\mathbf{v}$ is the velocity field, 
$p$ is the pressure field, and $\mathbf{r}$ stands for three-dimensional positional vector.
The above equation is further subjected to the incompressibility condition, 
$\nabla\cdot\mathbf{v}=0$.
From these equations, one can obtain a linear relation between the 
time-dependent force $F(t)$ acting on a hard sphere of radius $a$ and its 
time-dependent velocity $V(t)$.
In the Fourier domain, it can be represented as 
\begin{equation}
V(\omega)=\frac{1}{6\pi\eta[\omega] a}F(\omega),
\label{GSR}
\end{equation}
where we use a bilateral Fourier transform for 
$V(\omega)=\int_{-\infty}^{\infty} {\rm d}t\, V(t)e^{-i\omega t}$ and $F(\omega)$, 
while we employ a unilateral one for 
$\eta[\omega]=\int_{0}^{\infty} {\rm d}t\, \eta(t)e^{-i\omega t}$.
Equation~(\ref{GSR}) is the GSR that has been successfully used in active 
microrheology experiments~\cite{GSOMS}, and its mathematical validity was also 
discussed before~\cite{Schnurr97}.

Next, we briefly explain the three-sphere model for a minimum swimmer  
introduced by Najafi and Golestanian~\cite{Golestanian04,Golestanian08}. 
As schematically shown in Fig.~\ref{model}, this model consists of three spheres of 
the same radius $a$ that are connected by two arms of lengths $L_1(t)$ and $L_2(t)$ 
which undergo time-dependent motions.
Their explicit time dependencies will be given later.
If we define the velocity of each sphere along the swimmer axis as $V_i(t)$ 
with $i=1,2,3$, we have 
\begin{align}
\dot{L}_1(t)&=V_2(t)-V_1(t),
\label{L1dot}
\\
\dot{L}_2(t)&=V_3(t)-V_2(t),
\label{L2dot}
\end{align}
where $\dot{L}_1$ and $\dot{L}_2$ indicate the time derivatives.

Due to the hydrodynamic effect, each sphere exerts a force $F_i$  on the viscoelastic 
medium and experiences a force $-F_i$ from it.
To relate the forces and the velocities in the frequency domain, we use the GSR in 
Eq.~(\ref{GSR}) and the Oseen tensor in which the frequency-dependent viscosity 
$\eta[\omega]$ is used instead of a constant one~\cite{MW95,Mason00}.
Assuming that $a\ll L_1, L_2 $, we can write~\cite{Golestanian04,Golestanian08} 
\begin{widetext}
\begin{align}
\label{eq3}
V_1(\omega)&=\frac{F_1(\omega)}{6\pi\eta[\omega] a}+\frac{1}{4\pi\eta[\omega]}\frac{F_2(\omega) \ast L_1^{-1}(\omega)}{2\pi}+\frac{1}{4\pi\eta[\omega]}\frac{F_3(\omega) \ast (L_1+L_2)^{-1}(\omega)}{2\pi},\\
\label{eq4}
V_2(\omega)&=\frac{1}{4\pi\eta[\omega]} \frac{F_1(\omega) \ast L_1^{-1}(\omega)}{2\pi}+\frac{F_2(\omega)}{6\pi\eta[\omega] a}+\frac{1}{4\pi\eta[\omega] }\frac{F_3(\omega) \ast L_2^{-1}(\omega)}{2\pi},\\
\label{eq5}
V_3(\omega)&=\frac{1}{4\pi\eta[\omega] }\frac{F_1(\omega)\ast (L_1+L_2)^{-1}(\omega)}{2\pi}+\frac{1}{4\pi\eta[\omega] }\frac{F_2(\omega) \ast L_2^{-1}(\omega)}{2\pi}+\frac{F_3(\omega)}{6\pi\eta[\omega] a},
\end{align}
\end{widetext}
where we have used the bilateral Fourier transform such as 
$L_1^{-1}(\omega)=\int_{-\infty}^\infty {\rm d}t\,[L_1(t)]^{-1}e^{-i\omega t}$.
Furthermore, the convolution of two functions are generally defined by 
$g_1(\omega)\ast g_2(\omega)=\int _{-\infty}^\infty {\rm d}\omega'\, g_1(\omega-\omega')g_2(\omega')$
in the above equations.

As in the original study, we are interested in autonomous net locomotion of the swimmer, 
and there are no external forces acting on the spheres.
If the inertia of the surrounding fluid can be neglected, we have the following force balance 
condition 
\begin{equation}
\label{eq6}
F_1(t)+F_2(t)+F_3(t)=0.
\end{equation}

Since Eqs.~(\ref{eq3})--(\ref{eq5}) involve the convolutions in the frequency domain, 
we cannot solve these equations for arbitrary $L_1(t)$ and $L_2(t)$. 
Here we assume that the two arms undergo the following periodic motions: 
\begin{align}
\label{L1cos}
L_1(t)&=\ell+d_1\cos(\Omega t),
\\
\label{L2cos}
L_2(t)&=\ell+d_2\cos(\Omega t-\phi).
\end{align}
In the above, $\ell$ is a constant length, $d_1$ and $d_2$ are amplitudes of the oscillatory 
motions, $\Omega$ is a common arm frequency, and $\phi$ is a mismatch in phases 
between the two arms.
In the following analysis, we generally assume that $d_1, d_2 \ll \ell$. 
The \textit{time-reversal symmetry} of the arm motion is present when $\phi=0$ and $\pi$.
Furthermore, we characterize the \textit{structural symmetry} of the swimmer by $d_1$ and $d_2$, 
i.e., the structure is symmetric when $d_1=d_2$ while it is asymmetric when $d_1 \ne d_2$.

Since the arm frequency is $\Omega$,  we assume that the velocities 
and the forces of the three spheres can be generally written as 
\begin{align}
V_i(\omega)& =V_{i,0}\,\delta(\omega)\nonumber\\
&+\sum_{n=1}^\infty \left[V_{i,n}\,\delta(\omega+n\Omega)+V_{i,-n}\,\delta(\omega-n\Omega)\right],
\label{Vex}\\
F_i(\omega)& =F_{i,0}\,\delta(\omega)\nonumber\\
&+\sum_{n=1}^\infty \left[F_{i,n}\,\delta(\omega+n\Omega)+F_{i,-n}\,\delta(\omega-n\Omega)\right],
\label{Fex}
\end{align}
where $i=1,2,3$ for the three spheres. 
Substituting Eqs.~(\ref{Vex}) and (\ref{Fex}) into the six coupled equations 
(\ref{L1dot}), (\ref{L2dot}), (\ref{eq3}), (\ref{eq4}), (\ref{eq5}) and 
(\ref{eq6}),  we obtain a matrix equation with infinite dimensions.

Under the conditions $d_1, d_2 \ll \ell$ and $a \ll \ell$,  we are allowed to 
consider only $n=-1, 0, 1$ and further approximate as $F_{i,\pm2}\approx 0$.
Then we can solve for six unknown functions $V_i(\omega)$ and $F_i(\omega)$, 
and further calculate the total swimming velocity 
$V=(V_1+V_2+V_3)/3$. 
Up to the lowest order terms in $a$, the average swimming velocity over one 
cycle of motion becomes~\cite{supplement} 
\begin{equation}
\overline{V} \approx \frac{7d_1d_2a\Omega}{24\ell^2}\frac{\eta'[\Omega]}{\eta_0}
\sin\phi 
-\frac{5(d_1^2-d_2^2)a\Omega}{48\ell^2}\frac{\eta''[\Omega]}{\eta_0},
\label{barVve}
\end{equation}
where $\eta'[\Omega]$ and $\eta''[\Omega]$ are the real and the imaginary parts of the 
complex shear viscosity, respectively, and $\eta_0=\eta[\Omega \rightarrow 0]$ is a 
constant zero-frequency viscosity.

The first term in Eq.~(\ref{barVve}) can be regarded as a viscous contribution and is 
present only if the time-reversal symmetry of the swimmer motion is broken, i.e., $\phi \ne 0,\pi$.
The second term, on the other hand, corresponds to an elastic contribution, and exists only 
when the structural symmetry of the swimmer is broken, i.e., $d_1 \ne d_2$. 
If we were able to control $d_1$, $d_2$ and $\Omega$ of the swimmer,  we first obtain 
$\eta'[\Omega]$ by measuring $\overline{V}$ as a function of $\Omega$ by setting 
$d_1=d_2$. 
Then we differentiate between $d_1$ and $d_2$ to see the change in $\overline{V}$, 
which then yields $\eta''[\Omega]$.
The corresponding complex shear modulus is simply obtained by $G[\Omega]=i \Omega \eta[\Omega]$.
This is a new type of active microrheology that we propose in this Letter.

\begin{table}[t]
\label{table}
\begin{center}
\caption{Locomotion of a three-sphere swimmer in a viscoelastic medium and the relevant
rheological information.}
  \begin{tabular}{|c||c|c|c|c|c|c|c|c|} \hline
    medium & \multicolumn{4}{|c|}{viscous} & \multicolumn{4}{|c|}{viscoelastic}\\ \hline
    time-reversal symmetry &  \multicolumn{2}{|c|}{Y} & \multicolumn{2}{|c|}{N}& \multicolumn{2}{|c|}{Y} & \multicolumn{2}{|c|}{N}  \\ \hline
     structural symmetry &  Y & N &Y & N &Y & N &Y & N\\ \hline\hline
     swimmer motion &N&N&Y&Y&N&Y&Y&Y\\\hline
     rheological information &$-$&$-$&N&N&$-$&$\eta''$&$\eta'$&$\eta',\eta''$\\\hline
  \end{tabular}
  \label{gst}
  \end{center}
\end{table}

For a purely Newtonian fluid, namely, for a medium characterized by a constant viscosity, 
the second term in Eq.~(\ref{barVve}) vanishes, and the first term coincides with the
expression obtained by Golestanian and Ajdari~\cite{Golestanian08}.
It should be noticed here, however, that the velocity $\overline{V}$ in this case no 
longer depends on the constant viscosity (because $\eta'[\Omega]/\eta_0=1$) and 
we cannot measure it by looking at $\overline{V}$.
Equation (\ref{barVve}) also implies that the swimmer cannot move in a purely elastic medium 
for which we have $\eta_0 \rightarrow \infty$.
Importantly, due to the presence of the second term, Purcell's scallop theorem breaks down for a 
three-sphere swimmer 
in a viscoelastic medium. 
Namely, even if the time-reversal symmetry of the swimmer motion is not broken, i.e., 
$\phi= 0,\pi$, the present swimmer can still move in a viscoelastic medium due to 
the second term as long as its structural symmetry is broken, i.e., $d_1\ne d_2$. 
According to Eq.~(\ref{barVve}), the motion of a three-sphere swimmer in a viscoelastic medium 
and the relevant rheological information are summarized in Table~I.

\begin{figure}[tbh]
\begin{center}
\includegraphics[scale=0.3]{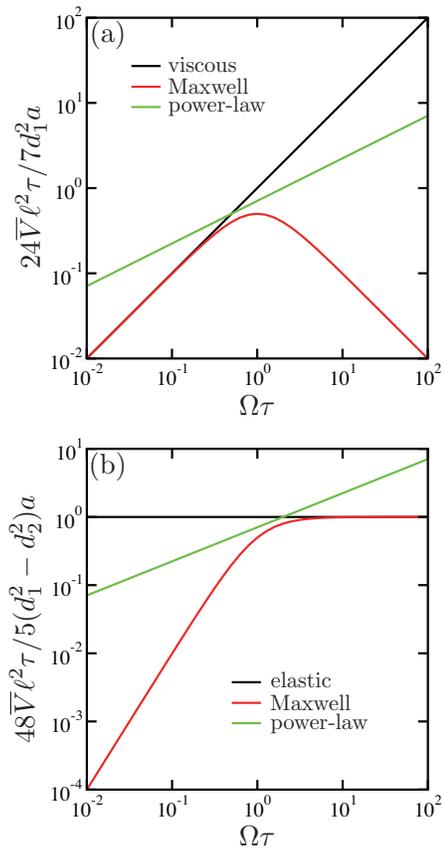}
\end{center}
\caption{
The average swimming velocity $\overline{V}$ as a function of $\Omega \tau$ 
where $\Omega$ is the arm frequency and $\tau$ represents 
either $\tau_{\rm M}$ for a Maxwell fluid (red lines) 
or $\tau_{\rm p}$ for a power-law fluid (green lines).
In the power-law model, we choose $\alpha=1/2$. 
(a) The viscous contribution by setting $\phi=\pi/2$ and $d_1=d_2$. 
Here $\overline{V}$ is scaled by $7d_1^2a/(24\ell^2\tau)$.
The case for a viscous fluid is plotted by the black line.
(b) The elastic contribution by setting $\phi=0$ and $d_1\ne d_2$.
Here $\overline{V}$ is scaled by $5(d_1^2-d_2^2)a/(48\ell^2\tau)$.
The case for an elastic medium is plotted by the black line.
}
\label{plotV}
\end{figure}

To further illustrate our result, we first assume that the surrounding viscoelastic medium 
is described by a simple Maxwell model.  
In this case, the frequency-dependent viscosity can be written as 
\begin{equation}
\eta[\omega]=\eta_0\frac{1-i\omega\tau_{\rm M}}{1+\omega^2\tau_{\rm M}^2}, 
\label{maxwell}
\end{equation}
where $\tau_{\rm M}$ is the characteristic time scale.
Within this model, the medium behaves as a viscous fluid for $\omega\tau_{\rm M} \ll 1$, 
while it becomes elastic for $\omega\tau_{\rm M} \gg 1$.
Using Eq.~(\ref{maxwell}), we can easily obtain the average swimming velocity in 
Eq.~(\ref{barVve}) as 
\begin{align}
\overline{V}& =\frac{7d_1d_2a\Omega}{24\ell^2}\frac{1}{1+\Omega^2\tau_{\rm M}^2}
\sin \phi \nonumber \\
& +\frac{5(d_1^2-d_2^2)a\Omega}{48\ell^2}
\frac{\Omega\tau_{\rm M}}{1+\Omega^2\tau_{\rm M}^2}.
\label{maxwellV}
\end{align}
The first viscous term increases as $\overline{V} \sim \Omega$ for $\Omega \tau_{\rm M} \ll 1$,
while it decreases as $\overline{V} \sim \Omega^{-1}$ for $\Omega \tau_{\rm M} \gg 1$.
This is a unique feature of the viscoelasticity~\cite{Lauga092,Fu07,Fu09},
but such a reduction occurs simply because the medium responds elastically in the 
high-frequency regime. 
On the other hand, the second elastic term increases as 
$\overline{V} \sim \Omega^2$ for $\Omega \tau_{\rm M} \ll 1$, and it approaches a constant 
for $\Omega \tau_{\rm M} \gg 1$.
In Fig.~\ref{plotV}(a), we plot the average swimming velocity $\overline{V}$ as a function 
of the dimensionless arm frequency $\Omega \tau_{\rm M}$ when 
$\phi=\pi/2$ and $d_1=d_2$. 
This plot corresponds to the first term in Eq.~(\ref{maxwellV}).
As a reference, the behavior of $\overline{V} \sim\Omega$ for a purely viscous fluid 
is also plotted.
Figure~\ref{plotV}(b) is a similar plot when $\phi=0$ and $d_1 \ne d_2$, and corresponds 
to the second term in Eq.~(\ref{maxwellV}).

As a different example, we next consider the case in which the viscoelastic medium is 
described by a power-law model such that~\cite{Granek11,Komura12,Komura15}
\begin{equation}
\eta[\omega]=G_0(i\omega)^{\alpha-1},
\label{power}
\end{equation}
where the exponent can take values $0 \le \alpha \le 1$.
With this expression, the complex shear modulus also obeys a  
power-law behavior, $G[\omega]=G_0(i\omega)^\alpha$.
The limits of $\alpha=0$ and $1$ correspond to the purely elastic and the purely viscous cases, 
respectively.
In the case of a power-law fluid, the average swimming velocity can be obtained from 
Eqs.~(\ref{barVve}) and (\ref{power}) as 
\begin{align}
\overline{V} & =\frac{7 d_1d_2a}{24 \ell^2\tau_{\rm p}}
(\Omega\tau_{\rm p})^\alpha\sin(\pi\alpha/2)\sin \phi \nonumber\\
&+\frac{5(d_1^2-d_2^2)a}{48\ell^2\tau_{\rm p}}
(\Omega\tau_{\rm p})^\alpha\cos(\pi\alpha/2),
\label{barVpow}
\end{align}
where $\tau_{\mathrm{p}}=(\eta_0/G_0)^{1/(1-\alpha)}$. 
Here we have assumed that the medium behaves as a purely viscous fluid in the 
low-frequency limit characterized by a finite viscosity $\eta_0$.
According to the above expression, the swimming velocity scales as 
$\overline{V} \sim \Omega^\alpha$ both in the first and the second terms.
For the purely viscous case of $\alpha=1$, the first term reduces to the result by 
Golestanian and Ajdari~\cite{Golestanian08}, while the second term vanishes. 
For the purely elastic case of $\alpha=0$, on the other hand, the first term vanishes
and the second term remains although the latter does not depend on the arm frequency 
$\Omega$ any more.
In Figs.~\ref{plotV}(a) and (b), we have also plotted the average velocity 
$\overline{V}$ as a function $\Omega \tau_{\rm p}$ when $\alpha=1/2$.
In both of these plots, the scaling behavior $\overline{V}\sim\Omega^{1/2}$ is seen.

Lauga considered an axisymmetric squirming motion of a sphere (squirmer) 
embedded in an Oldroyd-B fluid which typically represents a polymeric fluid~\cite{Lauga09}. 
He reported that the scallop theorem in a viscoelastic fluid breaks down if the squirmer
has a fore-aft asymmetry in its surface velocity distribution.
For a time-reversal deformation given by a simple sinusoidal gait, he showed 
that the average swimming velocity is given by 
$\overline{V} \sim\Omega {\rm De}/(1+{\rm De}^2)$, where the Deborah number
is given by ${\rm De}=\Omega \tau_{\rm O}$ with a characteristic relaxation time
$\tau_{\rm O}$ in the Oldroyd-B model.
Such a frequency dependence of the swimming velocity is identical to the second term 
of  Eq.~(\ref{maxwellV}) obtained for a Maxwell fluid although Eq.~(\ref{barVve}) 
is more general.
On the other hand, our result is different from that by Curtis and Gaffney~\cite{Curtis13},
because they showed that the swimming velocity in a viscoelastic medium is the same as 
that in a Newtonian fluid.

To summarize, we have proposed a new active microrheology using the 
Najafi-Golestanian's three-sphere swimmer.
The frequency dependence of the average swimming 
speed provides us with the complex shear viscosity of the surrounding 
viscoelastic medium. 
Here the viscous contribution can exist only when the time-reversal symmetry 
of the swimmer is broken, whereas the elastic contribution is present only if its structural 
symmetry is broken.

Even though the argument in this Letter is restricted to the artificial three-sphere swimmer, 
we expect that our basic concept can be still applied to more complex biological processes 
such as the motion of bacteria, the flagellated cellular swimming, or the beating of cilia. 
Since most of these phenomena take place in viscoelastic environment, we hope that the 
concept of our new active microrheology will be used in the future to reveal their mechanical 
and dynamical properties.

S.K. acknowledges support from the Grant-in-Aid for Scientific Research on
Innovative Areas ``\textit{Fluctuation and Structure}" (Grant No.\ 25103010) from the Ministry
of Education, Culture, Sports, Science, and Technology of Japan, and from 
the Grant-in-Aid for Scientific Research (C) (Grant No.\ 15K05250)
from the Japan Society for the Promotion of Science (JSPS).


\pagebreak

\onecolumngrid

\begin{center}
$\ $ 
{\large \textbf{Supplemental Materials: Swimmer-microrheology}}\\

$\ $ 

Kento Yasuda, Ryuichi Okamoto, and Shigeyuki Komura

\textit{Department of Chemistry, Graduate School of Science and Engineering,}

\textit{Tokyo Metropolitan University, Tokyo 192-0397, Japan}
\end{center}

\setcounter{equation}{0}
\renewcommand{\theequation}{S\arabic{equation}}   

\vspace{0.3in}

In this Supplemental Materials, we show the detailed derivation of Eq.~(13).
Substituting Eqs.~(9) and (11) into Eq.~(3), we obtain 
\begin{align}
&V_{2,0}-V_{1,0}=0, \\
&V_{2,1}-V_{1,1}=-i\pi d_1\Omega, \\
&V_{2,-1}-V_{1,-1}=i\pi d_1\Omega, \\
&V_{2,n}-V_{1,n}=0~~~\text{for $|n| \ge 2$}.
\label{eq1veT}
\end{align}
Similarly, substituting Eqs.~(10) and (11) into Eq.~(4), we obtain
\begin{align}
&V_{3,0}-V_{2,0}=0, \\
&V_{3,1}-V_{2,1}=\pi d_2\Phi_2\Omega, \\
&V_{3,-1}-V_{2,-1}=\pi d_2\Phi_1\Omega, \\
&V_{3,n}-V_{2,n}=0~~~\text{for $|n| \ge 2$}, 
\label{eq2veT}
\end{align}
where we have used the following notations
\begin{align}
&\Phi_1=i\cos\phi+\sin\phi, \\
&\Phi_2=-i\cos\phi+\sin\phi.
\end{align}

Next we expand Eqs.~(5), (6) and (7) in terms of the small quantities 
$d_1/\ell$ and $d_2/\ell$ while keeping only the lowest order terms. 
Substituting Eqs.~(11) and (12) into these three equations, we obtain
\begin{align}
V_{1,n}& \approx \frac{F_{1,n}}{6\pi\eta[-n\Omega] a}+\frac{1}{4\pi\eta[-n\Omega]\ell}\left(F_{2,n}-\frac{d_1F_{2,n+1}}{2\ell}-\frac{d_1F_{2,n-1}}{2\ell}\right)\nonumber\\
&+\frac{1}{4\pi\eta[-n\Omega]\ell}\left(\frac{F_{3,n}}{2}-\frac{d_1F_{3,n+1}}{8\ell}-\frac{d_1F_{3,n-1}}{8\ell}+\frac{id_2\Phi_1F_{3,n+1}}{8\ell}-\frac{id_2\Phi_2F_{3,n-1}}{8\ell}\right),
\label{eq3veT}\\
V_{2,n}& \approx \frac{1}{4\pi\eta[-n\Omega]\ell}\left(F_{1,n}-\frac{d_1F_{1,n+1}}{2\ell}-\frac{d_1F_{1,n-1}}{2\ell}\right)+\frac{F_{2,n}}{6\pi\eta[-n\Omega] a}\nonumber\\
&+\frac{1}{4\pi\eta[-n\Omega] \ell}\left(F_{3,n}+\frac{id_2\Phi_1F_{3,n+1}}{2\ell}-\frac{id_2\Phi_2F_{3,n-1}}{2\ell}\right),
\label{eq4veT}\\
V_{3,n}& \approx \frac{1}{4\pi\eta[-n\Omega]\ell }\left(\frac{F_{1,n}}{2}-\frac{d_1F_{1,n+1}}{8\ell}-\frac{d_1F_{1,n-1}}{8\ell}+\frac{id_2\Phi_1F_{1,n+1}}{8\ell}-\frac{id_2\Phi_2F_{1,n-1}}{8\ell}\right)\nonumber\\
&+\frac{1}{4\pi\eta[-n\Omega]\ell }\left(F_{2,n}+\frac{id_2\Phi_1F_{2,n+1}}{2\ell}-\frac{id_2\Phi_2F_{2,n-1}}{2\ell}\right)+\frac{F_{3,n}}{6\pi\eta[-n\Omega] a}.
\label{eq5veT}
\end{align}
Note that the couplings between different $n$-modes are involved in these equations.
Finally, substituting Eq.~(12) into Eq.~(8), we obtain
\begin{align}
\label{eq6veT}
&F_{1,n}+F_{2,n}+F_{3,n}=0.
\end{align}

The above set of equations constitute a matrix equation with infinite dimensions 
and cannot be solved in general.  
Under the assumption of $a\ll\ell$, however, we are allowed to consider only
$n=-1, 0, 1$ and further approximate as $F_{i,\pm2}\approx 0$. 
The justification of the latter approximation is also seen by solving 
Eqs.~(\ref{eq1veT}), (\ref{eq2veT}), (\ref{eq3veT}), (\ref{eq4veT}), 
(\ref{eq5veT}) and (\ref{eq6veT}) for $n=\pm2$ and taking the limit of $a\ll\ell$.
Hence the above set of equations can be solved for 18 unknowns, i.e., 
$V_{i,n}$ and $F_{i,n}$ for $i=1, 2, 3$ and $n=-1, 0, 1$.

The velocity of each sphere is simply obtained by the inverse Fourier transform,
$V_i(t)=(2\pi)^{-1} \int_{-\infty}^{\infty} {\rm d} \omega\, V_i(\omega) e^{i\omega t}$.
The average swimming velocity over one cycle of motion is then calculated by 
\begin{equation}
\label{barV}
\overline{V}=\frac{\Omega}{2\pi}\int_0^{2\pi/\Omega}{\rm d}t\, 
[V_1(t)+V_2(t)+V_3(t)]/3.
\end{equation}
Up to the lowest order terms in $a$, we finally obtain Eq.~(13).
In order to obtain more accurate higher order terms in $a$, one needs to 
take into account the higher order $n$-modes ($|n| \ge 2$).

\end{document}